\documentclass[twocolumn,aps,10pt,showkeys,showpacs,preprintnumbers,prd,superscriptaddress,nofootinbib]{revtex4-1}
\usepackage{graphicx}	
\usepackage{amssymb}
\usepackage{dcolumn}
\usepackage{mathtools}
\usepackage{amsmath}
\usepackage{xcolor}
\usepackage{color}
\usepackage[normalem]{ulem}
\usepackage{theorem}
\usepackage{subfigure, rotating, bm, array}
\usepackage[pagebackref=false, colorlinks=true]{hyperref}
\hypersetup{linkcolor=blue, citecolor=blue,urlcolor=blue}

\begin{document}
\title{Retrograde Precession of Relativistic Orbits and the Quest for Charged Black Holes}

\author{Parth Bambhaniya}
\email{grcollapse@gmail.com}
\affiliation{Instituto de Astronomia, Geofísica e Ciências Atmosféricas, Universidade de São Paulo, IAG, Rua do Matão 1225, CEP: 05508-090 São Paulo - SP - Brazil.}
\author{Meet J. Vyas}
\email{meet.v2@ahduni.edu.in}
\affiliation{International Centre for Space and Cosmology, Ahmedabad University, Ahmedabad 380009, Gujarat, India}
\author{Pankaj S. Joshi}
\email{psjcosmos@gmail.com}
\affiliation{International Centre for Space and Cosmology, Ahmedabad University, Ahmedabad 380009, Gujarat, India}
\author{Elisabete M. de Gouveia Dal Pino}
\email{dalpino@iag.usp.br}
\affiliation{Instituto de Astronomia, Geofísica e Ciências Atmosféricas, Universidade de São Paulo, IAG, Rua do Matão 1225, CEP: 05508-090 São Paulo - SP - Brazil.}
\date{\today}
\begin{abstract}

The S-stars around the center of the milky way galaxy provide us with detailed information about the nature of the supermassive compact object Sagittarius A* (Sgr A*). In this work, we derive the fully relativistic orbit equations for the case of the Reissner-Nordström (RN) and Kerr-Newman spacetimes. We solve these orbit equations numerically to analyze the periastron shift of relativistic orbits. We show that retrograde precession (or negative precession) of timelike bound orbits is possible in the case of naked singularity arising from these spacetimes. We have then compared our results with the non-charged Schwarzschild and Kerr spacetimes. This theoretical analysis of relativistic orbits would be helpful in either confirming or ruling out such charged black holes and naked singularities through the future trajectories of S-stars and will also help us constrain the geometry of Sgr A*.   

\vspace{.6cm}
$\boldsymbol{Key words}$ :  Retrograde precession, Relativistic orbits, Sgr A*, S-stars, Charged Black holes.
\end{abstract}
\maketitle

\section{Introduction}

Recently several groups such as the Event Horizon Telescope \cite{EHTC2022} (EHT), GRAVITY \cite{Abuter2020}, SINFONI \cite{Eisenhauer2005} and UCLA Galactic Center Group \cite{Ghez2005} have been trying to probe the nature of the supermassive compact object Sgr A* near the center of our galaxy. It has been approximated that the mass of Sgr A* is about $4.3\times10^6 M_{\odot}$,  where $M_{\odot}$ is the mass of the sun. It is located at a distance of $8.2$ kpc ($1~ $kpc $\sim$ $3\times 10^{16}$ km) from the Earth. There are many `S' stars (e.g. S2, S38, S102, etc.) which are orbiting around Sgr A*. These stars can be considered as massive particles following timelike bound orbits. The nature of the timelike bound orbits for different spacetimes has been widely discussed in various works \cite{Janis-Newman1965,Farina1993,Fujita2009, Hackmann2008,Hackmann2009,Hackmann2014,Pugliese2011,Pugliese2012,Pugliese2013,Dasgupta2012,Martinez2019,Potashov2019,Bhattacharya2020,Lin2021,Deng2020a,Deng2020b,Gao2020,Harada2023,Katsumata2024}.

The periastron shift of timelike bound orbits have been studied for Schwarzschild Black Hole, Joshi-Malafarina-Narayan-1 (JMN-1) and Joshi-Malafarina-Narayan-2 (JMN-2) in \cite{Bambhaniya2019a,Bambhaniya2019c}, and for Bertrand (BST) and Janis-Newman-Winicour (JNW) naked singularity spacetimes \cite{Bambhaniya2019b} also. A null naked singularity spacetime was studied \cite{Madan2024}, and the Kerr Spacetime \cite{Bambhaniya2021a}, Deformed Kerr Spacetime \cite{Bambhaniya2021b}, Rotating JNW Spacetime \cite{Bambhaniya2022}, and Rotating JMN Spacetime \cite{Bambhaniya2023} have also been investigated from such a perspective. Then the relativistic orbits of S2 star have also been studied using the astrometric and spectroscopic data \cite{Bambhaniya2024}. The detailed study and discussion on relativistic orbits are given in \cite{Bambhaniya2024b}. We have seen that for vacuum solutions of the Einstein field equations, namely the Schwarzschild and Kerr spacetimes, the bound trajectories of massive particles always precess in the direction of particle motion and therefore have a prograde periastron shift. 

According to the no-hair theorem \cite{Penrose1965}, black holes can be completely characterized with only three basic properties. These properties are their mass, charge, and spin. The effects of charge for the relativistic corrections to the bound timelike orbits of massive particles around charged spacetimes can help us to confirm or rule out charged geometries, using the orbits of S-stars around Sgr A*. This may lead to a more precise determination of the nature of the geometry of Sgr A*. Subsequently, various observational aspects of Sgr A* have been explored, including its shadow and accretion disk properties \cite{Vagnozzi:2022moj,Joshi:2020tlq,Saurabh:2022jjv,Saurabh:2023otl,Vertogradov:2024fva,Bambhaniya:2024hzb}, relativistic time delays in pulsar signals \cite{Rajwade:2016cto,Liu:2011ae,Psaltis:2015uza,Kalsariya:2024qyp}, tidal force effects \cite{Kesden:2011ee,Crispino:2016pnv,Lima:2020wcb,Arora:2023ltv}, energy extraction mechanisms \cite{Zaslavskii2012, Chen2022, Patel2023,Patel2022}, and more. Recently general relativistic hydrodynamical (GRMHD) simulation of accretion is performed for RN black hole and RN naked singularity \cite{Wlodek2024}. For the RN black hole, matter plunges into the horizon, similar to the Kerr case. While for the RN naked singularity, accreting matter forms a toroidal inner structure, generating powerful outflows. These properties would be useful for future observations performed by EHT.
Based on this reasoning, we consider the case of RN spacetime \cite{Reissner1916}, which is the spherically symmetric static charged solution to the Einstein equations, and Kerr-Newman spacetime \cite{Newman1965}, which is the axisymmetric and rotating charged solution to the same equations. 

In this work, we examine the fundamental properties of the RN and Kerr-Newman spacetimes. Subsequently, we derive the fully relativistic orbit equation for a test particle confined to the equatorial plane in both scenarios. The orbit equations are then numerically solved, and particle trajectories are analyzed for a specific set of fixed parameters to explore the occurrence of negative precession in these spacetimes. We have then compared our results with the Schwarzschild and Kerr spacetimes.

We have the following arrangement of the paper. In section (\ref{sec:two}), we review the basic properties of the RN and Kerr-Newman spacetimes. Subsequently, we derive the fully relativistic orbit equation for a test particle confined to the equatorial plane in both scenarios. 
In section (\ref{sec:three}), we use an approximate solution of the orbit equation and investigate the nature of periastron shift of the timelike bound orbits in the RN spacetime. In section (\ref{sec:five}), we carry out the same approximation analysis for the Kerr-Newman spacetime. In section (\ref{sec:six}), we discuss the results of this work. Throughout the paper, we use the units $G = c = 1$.
\section{Relativistic Orbit Equations in Charged Spacetimes}
\label{sec:two}

The static, spherically symmetric metric for RN spacetime is given as,
\begin{equation}
        \begin{split}
        ds^2 = & -\left(1 - \frac{2M}{r} + \frac{q^2}{r^2}\right) dt^2 
               + \left(1 - \frac{2M}{r} + \frac{q^2}{r^2}\right)^{-1} dr^2 \\
               &+ r^2 \left(d\theta^2 + \sin^2\theta \, d\phi^2\right) \,\, ,
        \end{split}
    \label{rn_metric}
\end{equation}
where, $M$ is the mass of the black hole, and $q$ is the charge. Similarly, the Kerr-Newman spacetime in Boyer-Lindquist coordinates is given as,
\begin{equation}
    \begin{split}
    ds^2 = & -\left(1-\frac{2M r}{\Sigma}\right)  dt^2 + \frac{\Sigma}{\Delta} dr^2 \\
    &+ \Sigma d\theta^2 + \left(r^2 + a^2 + \frac{2M r a^2 \sin^2\theta}{\Sigma}\right) \sin^2\theta d\phi^2 \\
    & - \frac{4 M r a \sin^2\theta}{\Sigma}  dt d\phi\,\, , 
    \end{split}
    \label{kn_metric}
\end{equation}
where $M$ and $q$ again represent the mass and charge of the black hole, $a$ is the dimensionless spin parameter, defined by $a = \frac{J}{M} $, where $J$ is the total angular momentum. We have also defined, $\Sigma = r^2 + a^2 \cos^2\theta$ and $\Delta = r^2 - 2Mr + a^2 + q^2$. The horizons are defined by the following condition on the space metric component  $g_{rr}\rightarrow\infty$, where we get the coordinate singularities. For the RN spacetime the relation leads to solving the following equation for obtaining the positions of the horizons.

\begin{equation}
    1-\frac{2M}{r} + \frac{q^2}{r^2} = 0 \,\,.
    \label{rn_horizon1}
\end{equation}
The solution to above equation is:
\begin{equation}
    r_\pm = M \pm \sqrt{M^2 - q^2}\,\,.
    \label{rn_horizon2}
\end{equation}
For the RN black hole, when $M > q$, there are two horizons: the event horizon at $r = r_+$ and the Cauchy horizon at $r = r_-$. In the extremal case, where $M = q$, these two horizons coincide at $r = M$, leading to an extremal RN black hole. However, no horizon forms when $M < q $ and the RN solution represents a timelike naked singularity.
Now for the Kerr-Newman spacetime, the positions of the horizons can be obtained by solving the following expression. 

\begin{equation}
    \Delta =  r^2 - 2Mr + a^2 + q^2 = 0\,\,.
    \label{kn_horizon1}
\end{equation}
The solution to above equation is:
\begin{equation}
    r_\pm = M \pm \sqrt{M^2 - a^2 - q^2},
    \label{kn_horizon2}
\end{equation}
Similarly, for the Kerr-Newman black hole, when $M^2 > a^2 + q^2 $, two horizons exist: the event horizon at $ r = r_+ $ and the Cauchy horizon at $ r = r_- $. In the extremal case, where $ M^2 = a^2 + q^2 $, the two horizons coincide, resulting in an extremal Kerr-Newman black hole. If $ M^2 < a^2 + q^2 $, the horizon fails to form, and the solution again describes a naked singularity. The RN and Kerr-Newman metrics are independent of $t$ and $\phi$, therefore, the conserved energy ($e$) and the angular momentum ($L$) per unit rest mass are associated with the temporal and rotational symmetries, given by the Killing vector fields $\zeta^{\nu}=(1,0,0,0)$ and $\eta^{\nu}=(0,0,0,1)$. Further, we restrict the test particle's orbit to the equatorial plane ($\theta = \frac{\pi}{2}$) for simplicity without loss of generality. 

\subsection{Orbit equation for RN spacetime}

\begin{figure}[h]
    \centering
    \includegraphics[width=1\linewidth]{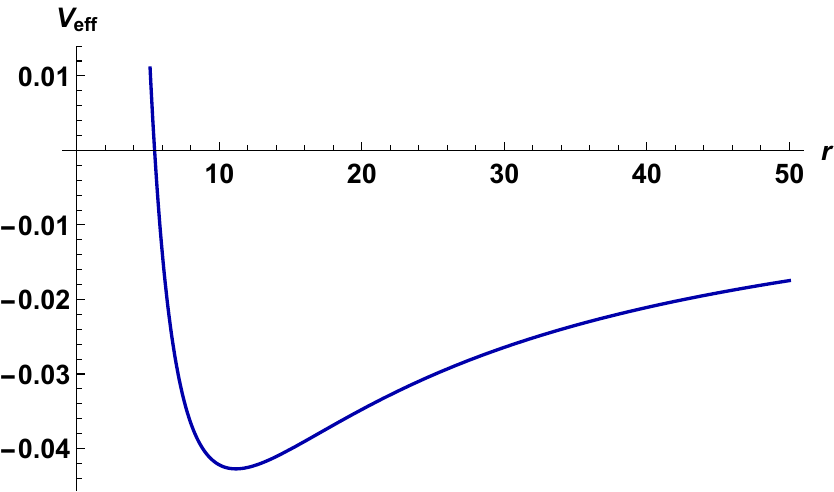}
    \caption{Effective potential plot of RN spacetime for $L=3$, $q=2$, $M=1$, $E=-0.02$, and $0 \leq r \leq 50$}
    \label{fig:veff_rn}
\end{figure}

We first write the expressions of $e$ and $L$ for the RN spacetime, which are given by,

\begin{equation}
    L =\eta _{\mu}U^{\mu}=g_{\mu \nu}\eta ^{\nu}U^{\mu}=g_{\varphi \varphi}U^{\varphi}
    \label{rn_conserved1},
\end{equation}
\begin{equation}
    e= -\zeta _{\mu}U^{\mu}=-g_{\mu \nu}\zeta ^{\nu}U^{\mu}=-g_{tt}U^t \,\,,
    \label{rn_conserved2}
\end{equation}
where $U^\mu$ are the components of the four-velocity of a test particle and $g_{tt} = -\left(1-\frac{2M}{r} + \frac{q^2}{r^2}\right) $, $g_{rr} = \left(1-\frac{2M}{r} + \frac{q^2}{r^2}\right)^{-1}$, $g_{\theta\theta} = g_{\phi\phi} = r^2$. Eqs. (\ref{rn_conserved1}), (\ref{rn_conserved2}) can be solved for $U^t$ and $U^{\phi}$, after which we get,

\begin{equation}
    U^t = \left(1-\frac{2M}{r} + \frac{q^2}{r^2}\right)^{-1} e \,\, ,
    \label{rn_four0}	
\end{equation}
\begin{equation}
    U^{\phi} = \frac{L}{r^2} \,\, .
    \label{rn_four1}
\end{equation}
Using the normalization condition $U^{\alpha}U_{\alpha}=-1$, for timelike geodesics and using Eqs. (\ref{rn_four0}), (\ref{rn_four1}), we can derive the radial component of the four-velocity as follows,	

\begin{equation}
U^{r} = \pm \frac{N_1 \cdot N_2}{r \sqrt{r (r - 2M) + q^2}} \,\, ,
\label{rn_four2}
\end{equation}
where, 
\begin{equation}
N_1 = \sqrt{\frac{-2 M r + q^2 + r^2}{r^2}} \,\, ,
\end{equation}

\begin{equation}
N_2 = \sqrt{r^2 \left(r \left(\left(e^2 - 1\right) r + 2 M\right) - q^2\right) - L^2 \left(r (r - 2 M) + q^2\right)} \,\, .
\end{equation}
Here, $\pm$ signatures correspond to radially outgoing and ingoing timelike geodesics respectively. The expression, in Eq.~(\ref{rn_four2}), is equivalent to the kinetic energy of a test particle. The total relativistic energy of the massive particle in any spacetime is defined as,

\begin{equation}
    E = \frac{1}{2} (e^2 - 1) = \frac{1}{2} (U^r)^2 + V_{eff}(r)\,\, .
    \label{rn_E1}
\end{equation}
Using the expression of $U^r$, Eq.~(\ref{rn_four2}) and the expression of total relativistic energy $E$, Eq.~(\ref{rn_E1}), we get the following {relation for} the  effective potential,
\begin{equation}
    V_{eff}(r) = \frac{L^2}{2 r^2} - \frac{M}{r} - \frac{M L^2}{r^3} + \frac{q^2}{2 r^2} + \frac{q^2 L^2}{2 r^4} \,\,.
    \label{rn_veff1}
\end{equation}
The above expression of the effective potential is only applicable for equatorial timelike geodesics. For timelike bound orbits, the total energy of the particle is greater than or equal to the minimum effective potential. The minimum effective potential is determined by,
\begin{equation}
    \frac{dV_{eff}}{dr}\bigg|_{r_b} = 0 \ ; \ \frac{d^2 V_{eff}}{dr^2}\bigg|_{r_b}>0\,\, ,
    \label{rn_con1}
\end{equation}
where the effective potential has a minimum at $r=r_b$. The minimum effective potential at $r = r_b$ is obtained by substituting $r_{b}$ in Eq. (\ref{rn_veff1}). The plot for $V_{eff}(r)$ is shown in fig. (\ref{fig:veff_rn}). 

\begin{figure}[h]
    \centering
    \includegraphics[width=1\linewidth]{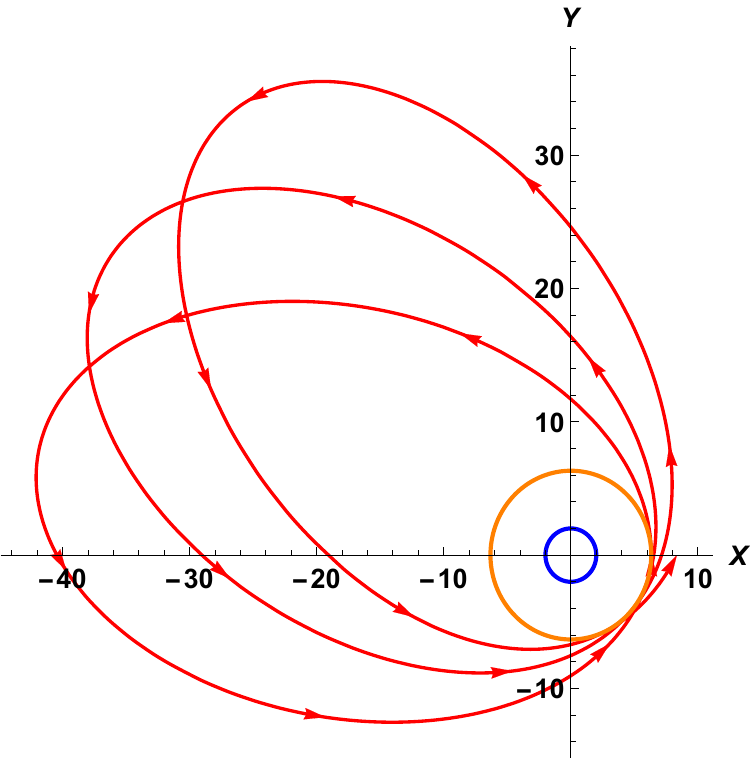}
    \caption{Relativistic orbits in the Reissner-Nordstrom Spacetime for $L=3$, $q=2$, $M=1$, $E=-0.02$, and $r_{min} = 6.3282$}
    \label{fig:orbit_rn}
\end{figure}

The timelike bound orbits exist for $V_{min}\leq E<0$. Using the bound orbit conditions, we can determine the shape of the orbits, which shows how $r$ changes in the equatorial plane with respect to $\phi$,
\begin{equation}
\frac{dr}{d\phi} = \frac{r \cdot N_1 \cdot N_2}{L \sqrt{r (r - 2 M) + q^2}} \,\, ,
\label{rn_radial_1}
\end{equation}
where, 
\begin{equation}
N_1 = \sqrt{\frac{-2 M r + q^2 + r^2}{r^2}} \,\, ,
\end{equation}

\begin{equation}
N_2 = \sqrt{r^2 \left(r \left(\left(e^2 - 1\right) r + 2 M\right) - q^2\right) - L^2 \left(r (r - 2 M) + q^2\right)}\,\, .
\end{equation}
Using Eq.~(\ref{rn_radial_1}), we can derive the second-order differential equation, the solutions of which represent the orbits of a massive test particle in RN spacetime,

\begin{align}
    \frac{d^2u}{d\phi^2} = & \frac{
    M - u(\phi) \left(L^2 + q^2 + L^2 u(\phi) \left(-3 M + 2 q^2 u(\phi)\right)\right)
    }{L^2} \, ,
    \label{rnorbit1}
\end{align}
where we have defined $u(\phi) = \frac{1}{r}$. We numerically solve the above orbit Eq.~(\ref{rnorbit1}) to investigate the nature and shape of timelike bound orbits of a freely falling test particle in RN naked singularity spacetime. Fig.~(\ref{fig:orbit_rn}) shows the timelike bound orbits of a test particle in this case. In fig.~(\ref{fig:orbit_rn}), we consider the total energy of the particle $E=-0.02$, the angular momentum $L=3$, the charge $q=2$, and the mass of the black hole $M=1$. The orbits shown by solid red lines correspond to the numerical solutions of the orbit equation, the orange circle shows the minimum approach of a particle towards the center, and the blue circle represents the event horizon at $r_+$. The orbits in fig.~(\ref{fig:orbit_rn}) show a negative precession as it shifts in the opposite direction of the particle's orbiting direction. 

\subsection{Orbit equation for Kerr-Newman spacetime}
\begin{figure}[ht]
    \centering
    \includegraphics[width=1\linewidth]{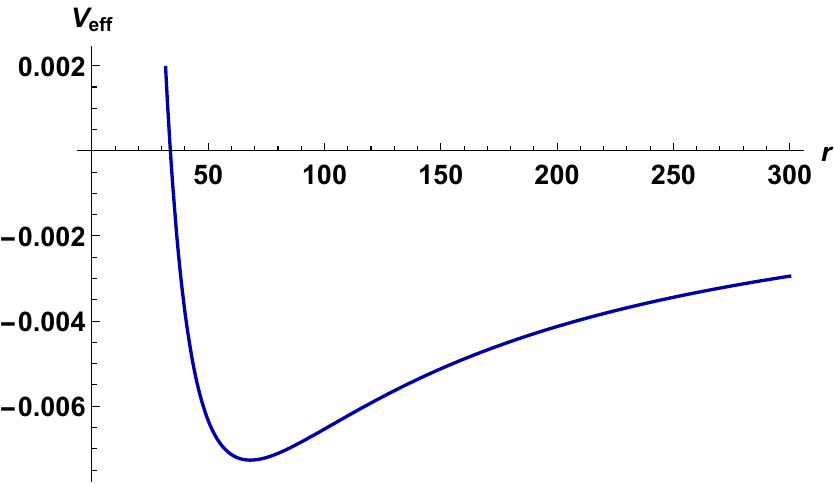}
    \caption{ Effective potential of Kerr-Newman spacetime}
    \label{fig:veff_kn}
\end{figure}

Similarly, we now write the conserved energy ($e$) and the angular momentum per unit of rest mass ($L$) for the Kerr-Newman spacetime, which are given as,

\begin{equation}
     L =\eta _{\mu}U^{\mu} = -g_{t\phi} U^t + g_{\phi\phi} U^{\phi}\,\, ,
     \label{kn_conserved2}
\end{equation}

\begin{equation}
    e= -\zeta _{\mu}U^{\mu}=g_{tt} U^t + g_{t\phi} U^{\phi}\,\, ,
    \label{kn_conserved1}
\end{equation}
where $U^\mu$ are again the components of the four velocity of a test particle and $g_{tt} = \left(1-\frac{2M r}{\Sigma}\right) $, $g_{rr} = \frac{\Sigma}{\Delta}$, $g_{\theta\theta} = \Sigma$, $g_{\phi\phi} = \left(r^2 + a^2 + \frac{2M r a^2 }{\Sigma}\right)$, and $g_{t\phi} = \frac{2M r a }{\Sigma} $. Using Eqs. (\ref{kn_conserved2}) and (\ref{kn_conserved1}), we can solve for $U^t$ and $U^{\phi}$:

\begin{equation}
    U^t =  \frac{1}{r^2-2Mr + a^2} \left[\left(r^2 + a^2 + \frac{2M  a^2 }{r}\right) e - \left(\frac{2M  a }{r}\right) L\right] \,\, ,
    \label{kn_four0}	
\end{equation}

\begin{equation}
    U^{\phi} = \frac{1}{r^2-2Mr + a^2} \left[\left(\frac{2M a }{r}\right) e + \left(1 - \frac{2M }{r}\right) L\right]\,\, .
    \label{kn_four1}
\end{equation}
Using the normalization condition again $U^{\alpha}U_{\alpha} = -1$, for timelike geodesic and Eqs. (\ref{kn_four0}) and (\ref{kn_four1}), we derive the $r$-component of the four-velocity $U^r$ as:

\begin{equation}
U^{r} = \sqrt{N_2} \,\, ,
\label{kn_four2}
\end{equation}
where, 
\begin{equation}
N_2 = -\frac{N_1}{r^4} + e^2 - 1 \,\, ,
\end{equation}
and
\begin{equation}
N_1 = N_{1a} + N_{1b} + N_{1c} \,\, ,
\end{equation}
given by,
\begin{equation}
N_{1a} = r^2 \left(a^2 \left(-\left(e^2-1\right)\right) + L^2 + q^2\right) \,\, ,
\end{equation}

\begin{equation}
N_{1b} = - 2 M r (L - a e)^2 \,\, ,
\end{equation}

\begin{equation}
N_{1c} = q^2 (L - a e)^2 - 2 M r^3 \,\, .
\end{equation}
where the $\pm$ signs again correspond to radially outgoing and ingoing timelike geodesics, respectively. The expression in Eq. (\ref{kn_four2}) is analogous to the kinetic energy of a test particle. The total relativistic energy is the same for the massive particle as in Eq. (\ref{rn_E1}). Using the expression of $U^r$ Eq. (\ref{kn_four2}) and the expression for the total relativistic energy Eq. (\ref{rn_E1}), we get the effective potential:

\begin{equation}
V_{eff}(r) = \frac{N_1 + N_2}{2 r^4} \,\, ,
\label{kn_veff1}
\end{equation}
where,
\begin{equation}
N_1 = (-a e + L)^2 q^2 - 2 (-a e + L)^2 M r \,\, ,
\end{equation}
and
\begin{equation}
N_2 = (-a^2 (-1 + e^2) + L^2 + q^2) r^2 - 2 M r^3 \,\, . 
\end{equation}
Applying the same reasoning as in the RN spacetime, where the condition that the particle's total energy must be greater than or equal to the minimum effective potential was used, we impose similar criteria for bound orbits as stated in Eq. (\ref{rn_con1}), with the effective potential reaching a minimum at \(r = r_b\). The minimum effective potential at \(r = r_b\) is determined by substituting \(r_b\) into Eq. (\ref{kn_veff1}) and the plot of \(V_{eff}(r)\) is shown in fig. (\ref{fig:veff_kn}). 

\begin{figure}[ht]
    \centering
    \includegraphics[width=1\linewidth]{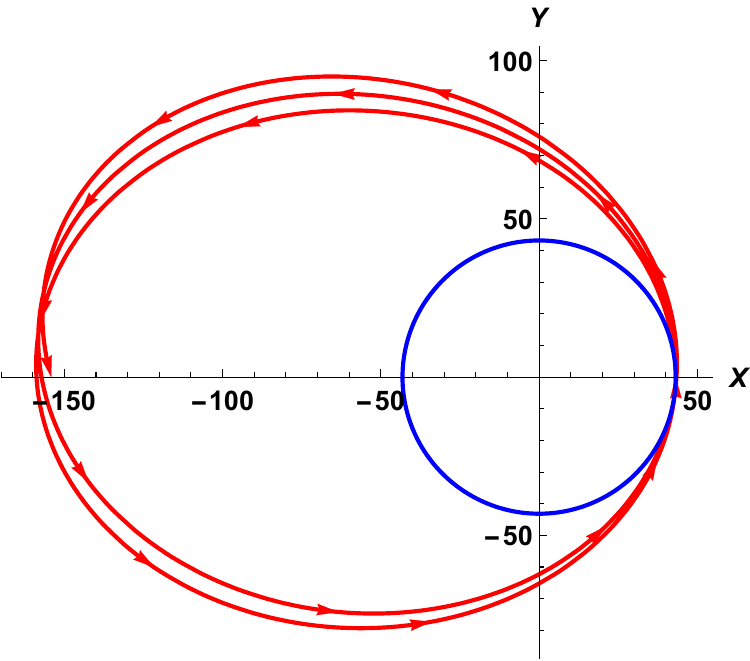}
    \caption{Relativistic orbits in the Kerr-Newman Spacetime for $L=8$, $q=2.5$, $M=1$, $E=-0.005$, $a = 0.8$, and $0 \leq r \leq 300$}
    \label{fig:orbit_kn}
\end{figure}

The timelike bound orbits exist for $V_{min}\leq E<0$. Using the bound orbit conditions, we can determine the shape and nature of the orbits, which gives how $r$ changes in the equatorial plane with respect to $\phi$, which is given as 


\begin{equation}
        \frac{dr}{d\phi} = \frac{U^{r}}{U^{\phi}} \,\, ,
    \label{kn_shapeorbits}
\end{equation}
using Eq.~(\ref{kn_shapeorbits}), we can derive a second-order differential orbit equation of a massive test particle in Kerr-Newman spacetime, which is given as 

\begin{equation}
\begin{aligned}
    \frac{d^2u}{d\phi^2} = {} & A + B + C + D \,\,,
    \label{knorbit}
\end{aligned}
\end{equation}
where $A$,$B$,$C$, and $D$ are all functions of $M,a,q,L,$ and $u(\phi)$. The complete functions are defined in the Appendix (\ref{sec:full_equations}). We again define for this case $u(\phi) = \frac{1}{r}$. We numerically solve the above orbit equation given by Eq.~(\ref{knorbit}) to investigate the nature and shape of timelike bound orbits of a test particle that is freely falling in the Kerr-Newman spacetime. Fig.~(\ref{fig:orbit_kn}) shows the bound orbits of a test particle in the Kerr-Newman naked singularity spacetime. In fig.~(\ref{fig:orbit_kn}), we consider the total energy of the particle $E = -0.005$, the angular momentum $L=8$, the charge $q=2.8$, the spin $a=0.8$, and the mass of the black hole to be $M=1$. The orbit shown by solid red lines corresponds to the numerical solutions to the orbit equation. The blue circle represents a minimum approach towards the center. The orbits in fig.~(\ref{fig:orbit_kn}) show a negative precession as angular distance traveled by a test particle is less than $2\pi$ or subsequently orbit will shifts in the opposite direction of the orbiting particle's direction.  

\section{An approximate solution of the Orbit Equation for the RN spacetime}
\label{sec:three}

In this section, we investigate whether a positive or negative precession of timelike orbits of a particle is possible for any value of $L$, $e$, and $q$. To achieve the same, we use an approximate method where we only consider low-eccentricity orbits. Therefore, for the approximate solution of the Eq.~(\ref{rnorbit1}), we only consider terms that are up to the first order in the eccentricity $\epsilon$. This method is extensively discussed in \cite{Bambhaniya2019a, Struck2006,Struck2015}. The approximate solution can be written as, 

\begin{equation}
u(\phi)=\frac{1}{M p}(1+\epsilon\cos(m\phi)+\mathcal{O}(\epsilon^2))\,\, ,
\label{loweccentricorbit}
\end{equation}
where $p$ and $m$ are real positive constants and $m>1$ represents the precession of the timelike bound orbits in the negative direction, while $m<1$ represents the precession in the positive direction. This comes from the fact that when $m<1$, the particle has to travel angular distance of more than $2 \pi$, starting from a periastron point to the next one, resulting in positive precession of the orbits. In contrast, a value of $m > 1$ corresponds to the case where the particle has to travel angular distance of less than $2 \pi$ to reach the next successive periastron point, resulting in a negative precession of the orbits. Substituting the expression for \(u(\phi)\) into the orbit Eq. (\ref{rnorbit1}), we separate the zeroth-order and first-order terms in \(\epsilon\). This process yields an expression for \(m\) in terms of \(p\). From the zeroth-order terms, a cubic polynomial in \(p\) is obtained,

\begin{equation}
    M^2 p^2 (-M^2 p + q^2) + L^2 (M^2 (-3 + p) p + 2 q^2) = 0.
    \label{polynomial}
\end{equation}
It is quite difficult to solve the cubic equation analytically, therefore we solve it numerically instead and get the values of $p$. We obtain the following expression for $m$ from the first order term in eccentricity $(\epsilon)$, 

\begin{equation}
m^2 = 1 + \frac{q^2}{L^2} + \frac{6 (-p + \frac{q^2}{M^2})}{p^2}.
\label{m_plot_eq_rn}
\end{equation}
Now, substituting numerical solutions of $p$ into Eq. (\ref{m_plot_eq_rn}), we get the numerical values and the plot for $m$. As we have mentioned before, $m>1$ implies negative precession of the timelike bound orbits in the RN spacetime. Therefore, we verify whether there exist any parameter space regions where $m$ is greater than one. 

\begin{figure}[ht]
    \centering
    \includegraphics[width=1\linewidth]{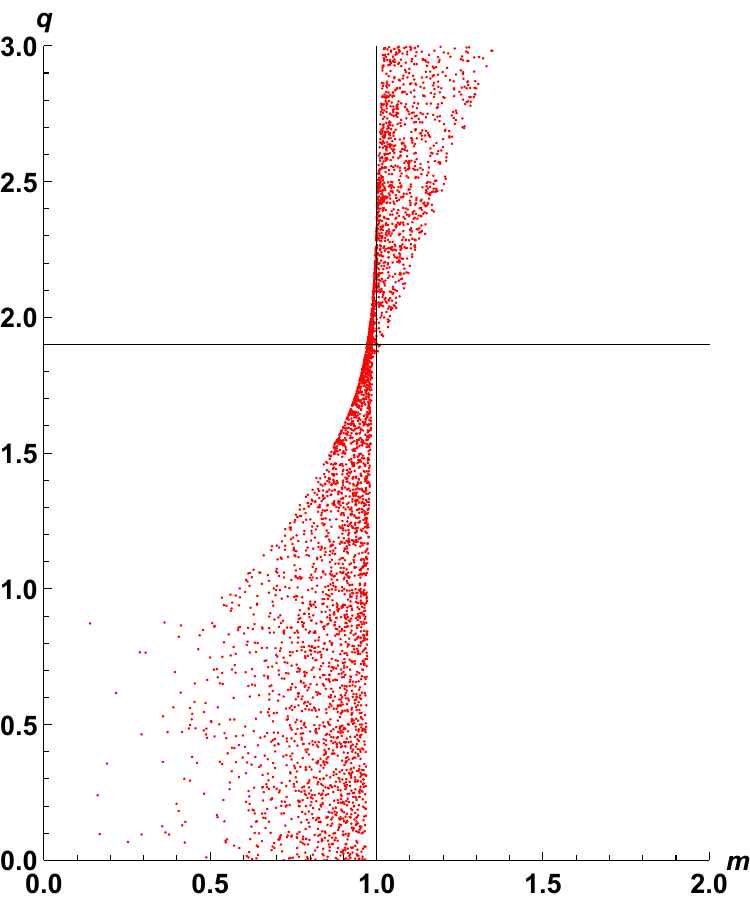}
    \caption{ $m-q$ plot for the RN spacetime solution Eq. (\ref{m_plot_eq_rn})  for the parametric range: $M=1$, $~3\leq L\leq 10$, $~0.965\leq e\leq 0.999$. }
    \label{fig:mplot_rn}
\end{figure}

To derive numerical solutions of $p$, we consider some specific physically realistic parameter ranges for the parameters $q,L$ and $e$. In our numerical analysis, we always consider $M=1$, $q$ varies from $0$ to $3$, $L$ varies from $3$ to $10$, and $e$ varies from $0.965$ to $0.999$. The charge parameter $q>3$ is not physically realistic as it may represent somewhat extreme charge situations. When the specific or conserved energy is $e<0.96$, solutions for bound stable orbits are difficult to construct. Moreover, the specific energy cannot have values beyond one, since the total energy of the particle becomes positive, which implies unbound orbits in the RN and Kerr-Newman spacetimes. Our approximate solution mentioned in Eq.~(\ref{loweccentricorbit}) is not a good approximation for the orbits of high eccentricity. For $L$ less than $3$, the eccentricity of the orbits becomes very high, and therefore, using our approximation method, we cannot proceed with numerical analysis in the $L<3$ region. However, the low eccentricity orbit approximation shows the presence of retrograde precession in RN naked singularity as shown in fig. (\ref{fig:mplot_rn}).

We know that $m$ is a function of $q,L, \text{and} \ e$ which allows for multiple values of $m$ for one value of $q$. We also remark that below $q<1$, we have the RN black hole spacetime; at $q=1$, we have the case of the extremal RN spacetime. In both of these cases we observe positive precession of the timelike bound orbits. For the case of $q>1$, we have the RN naked singularity spacetime, where we can see that for $q>1.9$, we have $m>1$, which corresponds to negative precession of orbits. In the next section, we will do similar analysis for Kerr-Newman spacetime. 
    
\section{An approximate solution of the Orbit Equation for the Kerr-Newman spacetime}
\label{sec:five}
In this section, we analyze the presence of negative or retrograde precession for the Kerr-Newman spacetime for any value of $L,e,q$, and $a$. We again take the low eccentric orbit approximation given by Eq. (\ref{loweccentricorbit}). We then substitute the expression for $u(\phi)$ in Eq. (\ref{knorbit}) and then separate the zeroth and first order terms in eccentricity $\epsilon$. By doing this we again obtain an equation for $m$ in terms of $p$. By separating the zeroth order terms, we get the equation of $p$, which is given as 

\begin{eqnarray}
    &g_7&(L,a,e,M,q) p^7 + g_6(L,a,e,M, q) p^6 \nonumber \\
    &+& g_5(L,a,e,M,q) p^5 + g_4(L,a,e,M, q) p^4 \nonumber \\
    &+& g_3(L,a,e,M,q) p^3 + g_2(L,a,e,M, q) p^2  \nonumber\\
    &+& g_1(L,a,e,M, q) p  + g_0(L,a,e,M,q) = 0 .\,\,  
    \label{kn_p_polynomial}
\end{eqnarray}
It is very difficult to solve the high-order equation in $p$ analytically and therefore, we have solved it numerically. We now separate the first-order terms and get the expression of $m^2$, which is given below.
\begin{eqnarray}
    m^2 &=& \frac{1}{p^2 \left(\left(2 p-Q^2\right) (a k-h)+h p^2\right)^4} (( f_{10}(L, a, e, M, q) p^{10} \nonumber \\
    &+& f_9(L, a, e, M, q) p^9 + f_8(L, a, e, M, q) p^8 \nonumber \\
    &+& f_7(L, a, e, M, q) p^7 + f_6(L, a, e, M, q) p^6 \nonumber \\
    &+& f_5(L, a, e, M, q) p^5 + f_4(L, a, e, M, q) p^4  \nonumber \\
    &+& f_3(L, a, e, M, q) p^3 + f_2(L, a, e, M, q) p^2  \nonumber \\
    &+& f_1(L, a, e, M, q) p + f_0(L, a, e, M, q)) = 0 \,\, . \nonumber\\ 
    \label{kn_m_polynomial}
\end{eqnarray}
The complete functions for the zeroth and first order terms are mentioned in the Appendix (\ref{sec:full_equations}). We then take the parameter space to check for the precession of the orbits in the low eccentricity approximation. As in the previous case, the range of $L < 3$ is not possible as the low eccentricity approximation does not hold and we may obtain highly eccentric orbits.
When the specific energy \(e\) drops below 0.965, finding solutions for bound stable orbits becomes highly challenging. Additionally, values of \(e > 1\) are not permissible, as the particle's total energy would become positive, corresponding to unbound orbits.
We take the values of $a$ between $-1$ and $1$ in accordance with the previous study \cite{Bambhaniya2021a} which gives a good approximation of the spin parameter space. 

The values of $q > 3$ are not considered realistic as they represent extreme charge situations and can be avoided. Similarly we can say that for $q < 0.5$ the effect of charge is not very evident and therefore we span the parameter space taking value of $q = (0.5,3)$. Therefore, we have considered the values of $L$ is from $3$ to $10$, $e$ is from $0.965$ to $0.999$, $a$ is from $-1$ to $1$ while keeping the values of $q$ in two different cases as $0.5$ and $3$. 
 
\begin{figure}[b]
    \centering
    \includegraphics[width=0.45\textwidth]{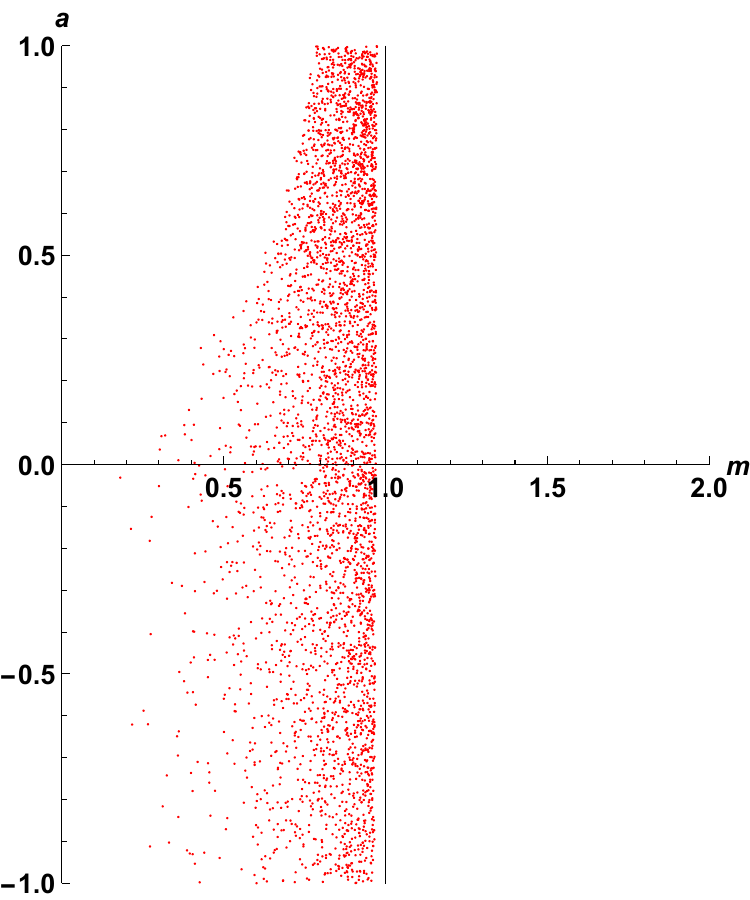}
    \caption{$m-a$ plot for the Kerr-Newman spacetime with parameters $M=1$, $3 \leq L \leq 10$, $0.965 \leq e \leq 0.999$, $-1 \leq a \leq 1$, $q=0.5$. For this plot, $0 < m < 1$, which represents positive precession of timelike bound orbits.}
    \label{fig:m_plot_case-1_kn}
\end{figure}
\begin{figure}[ht]
    \centering
    \includegraphics[width=0.45\textwidth]{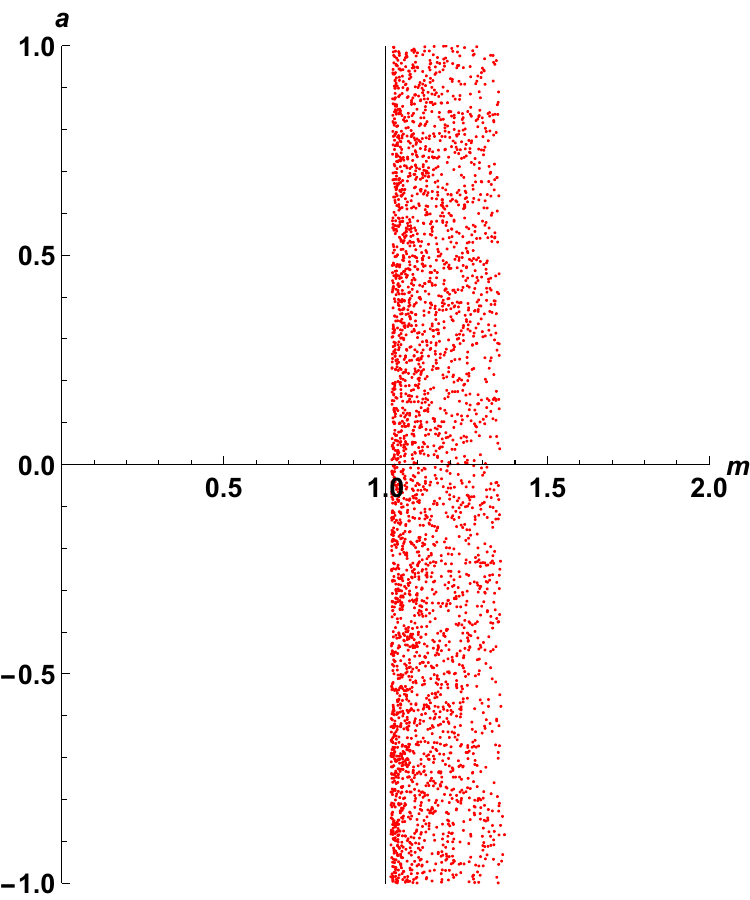}
    \caption{$m-a$ plot for the Kerr-Newman spacetime
       with parameters $M=1$, $3 \leq L \leq 10$, $0.965 \leq e \leq 0.999$, $-1 \leq a \leq 1$, $q=3$. For this plot, $m > 1$, which represents negative precession of timelike bound orbits.}
    \label{fig:m_plot_case-2_kn}
\end{figure}
From figs. (\ref{fig:m_plot_case-1_kn}) and (\ref{fig:m_plot_case-2_kn}), we infer that negative precession of particles is possible in the Kerr-Newman naked singularity spacetime. For lower values of charge, we still get positive precession of orbits, while for higher values of charge, we observe a negative precession of orbits. Therefore, negative precession of timelike bound orbits is also possible in the Kerr-Newman naked singularity spacetime.

\subsection{Periastron shift comparison with non-charged spacetimes}
\begin{figure}[t]
    \centering
    \includegraphics[width=0.89\linewidth]{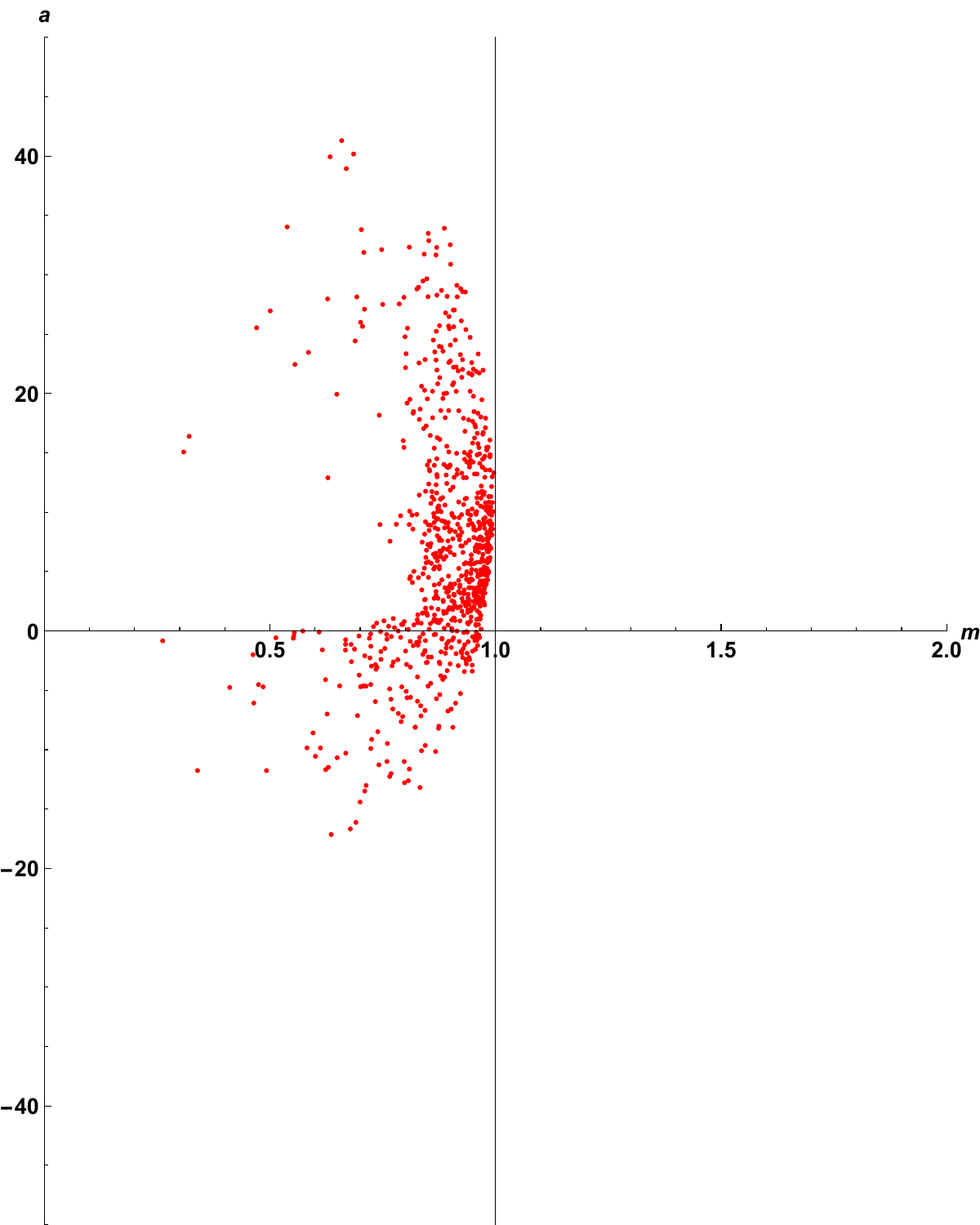}
    \caption{ $m-a$ plot for the Kerr spacetime with parameters $M=1$, $3 \leq L \leq 10$, $0.965 \leq e \leq 0.999$, $-1 \leq a \leq 1$, $q=0$. For this plot, $0 < m < 1$, which represents positive precession of timelike bound orbits.}
    \label{fig:kerr_m_plot}
\end{figure}

When we compare charged and non-charged rotating spacetimes for the nature of relativistic orbits, we find that non-charged Kerr spacetime with $q=o$, does not show retrograde precession for valid parameter ranges as defined above. One can see from the $m$ plot of Kerr spacetime (see fig. (\ref{fig:kerr_m_plot})), where we have $m<1$, which corresponds to positive or prograde precession of relativistic orbits. Therefore, the effect of charge on the spacetime geometry will be observed from the relativistic orbits in the Kerr-Newman spacetime. Similarly, we can find the effect of charge on the nature of relativistic orbits when we compare RN and Schwarzschild spacetimes. The $m-q$ plot is given for RN spacetime in fig. (\ref{fig:mplot_rn}), where we have $m<1$ and $m>1$, which correspond to prograde and retrograde precession of relativistic orbits respectively. Therefore, retrograde or negative precession of timelike bound orbits observed in the RN and Kerr-Newman naked singularity spacetimes can be considered as an effect of the presence of charge. In contrast, this type of orbital precession is not possible in the uncharged solutions such as Schwarzschild and Kerr spacetimes \cite{Bambhaniya2019a, Bambhaniya2021a}.

\section{Discussions and Conclusions}
\label{sec:six}
Previous studies have shown that the periastron precession of timelike bound geodesics is positive in Schwarzschild, Kerr, and Bertrand spacetimes \cite{Bambhaniya2019b,Bambhaniya2021a}. However, in JMN-1 spacetime, we have shown that the negative precession occurs for $0<M_0 < 1/3$, while for $1/3<M_0 <4/5$, we get positive precession \cite{Bambhaniya2019a}. On the other hand, for rotating JMN-1 spacetime, the range of $M_0$ for which we get negative or positive precession would depend upon the values of the spin parameter $a$ \cite{Bambhaniya2019a,Bambhaniya2024}. Similarly, negative precession occurs in static JNW spacetime for $0<n<0.5$ \cite{Bambhaniya2019b}, while for rotating JNW naked singularity spacetime,  the range of $n$ for which we get negative or positive precession would again depend upon the value of the spin parameter $a$ \cite{Bambhaniya2022}. Negative precession of orbits is also observed in the deformed Kerr spacetime for a given range of deformation parameter \cite{Bambhaniya2021b}. The key findings of the present study can be summarized as follows: 

\begin{itemize}

\item We derive the orbit equations for the RN and Kerr-Newman spacetimes and then solved numerically. Fig. (\ref{fig:orbit_rn}) illustrates the trajectories of a test particle in the RN naked singularity spacetime with the total mass \(M = 1\), specific angular momentum \(L = 3\), charge \(q = 2\), and total energy \(E = -0.02\). In this case, negative precession is observed, with the particle trajectory confined to the equatorial plane (\(\theta = \pi/2\)). Similarly, Figure (\ref{fig:orbit_kn}) displays the trajectories of a test particle in the Kerr-Newman naked singularity spacetime, considering the total mass \(M = 1\), specific angular momentum \(L = 8\), charge \(q = 2.5\), spin parameter \(a = 0.8\), and total energy \(E = -0.005\). Negative precession is again observed, with the particle trajectory restricted to the equatorial plane (\(\theta = \pi/2\)).
\item In section (\ref{sec:three}), we analyze an approximate solution to the orbit equation given by the Eq. (\ref{loweccentricorbit}), focusing on the periastron shift of timelike bound orbits in RN spacetime. This analysis is conducted under the assumption of small eccentricity ($\epsilon$), where we neglect second and higher-order terms of $\epsilon$. Using this approximation, we demonstrate that negative precession of timelike bound geodesics can occur in the RN naked singularity spacetime.
\item In section (\ref{sec:five}), we extend the low eccentricity approximation to the Kerr-Newman spacetime. Similar to the RN case, we show that negative precession is also possible for timelike bound orbits in the Kerr-Newman naked singularity spacetime under this approximation. Furthermore, we then draw a comparative analysis with the non-charged Black Holes. We find that retrograde or negative precession of timelike bound orbits observed in the RN and Kerr-Newman naked singularity spacetimes can be considered an effect of the presence of charges. In contrast, this type of orbital precession is not possible in the Schwarzschild and Kerr spacetimes as we have noted earlier.
\item As discussed in the introduction, S-stars can be modeled as massive particles orbiting the supermassive compact object at the center of our galaxy. Their orbits are continuously monitored using highly sensitive infrared instruments, including GRAVITY \cite{Abuter2018a}, SINFONI, and NACO, operated by the European Southern Observatory. A comprehensive dataset spanning 23 years of astrometric observations of the S2 star has been released \cite{Do2019,Abuter2018b,Hees2017}.
\item Recent studies have suggested that this intriguing feature of the retrograde periastron shift may be attributed to local energy density effects around the compact object, arising from the matter density near the star \cite{Igata2023}. It is proposed that this retrograde shift results from dominant local-density effects near the outer boundary of the matter distribution, rather than being caused by exotic spacetimes such as naked singularities. Furthermore, it has been argued that when the total mass of the system is fixed and the matter distribution is concentrated in a narrow region, local-density effects then tend to overshadow general relativistic effects, leading to retrograde precession. On the other hand, a more diffuse matter distribution favors prograde precession.
\item This interpretation, combined with our current analysis, suggests that the retrograde periastron precession observed in the naked singularity cases of RN and Kerr-Newman spacetimes can be understood as a consequence of the increased energy density of the electromagnetic field due to the higher charge in the naked singularity regime where \(q > M\). Consequently, the detection of negative precession in the trajectories of S-stars could provide valuable insights into the nature of matter distribution, spacetime geometry, and the nature of Sgr A*, contributing to a deeper understanding of our galaxy and the broader universe.

\end{itemize}

\acknowledgments{P. Bambhaniya and Elisabete M. de Gouveia Dal Pino acknowledge support from the São Paulo State Funding Agency FAPESP (grant 2024/09383-4). Elisabete M. de Gouveia Dal Pino also acknowledges the support from the São Paulo State Funding Agency FAPESP (grant 2021/02120-0) and CNPq (grant 308643/2017-8).}
\onecolumngrid

\appendix
\section{Full Length Equations}
\label{sec:full_equations}
\subsection{Kerr-Newman Orbit Equation}

Here, we describe the complete orbital equation as derived from Eq. (\ref{knorbit}), each variable $A$, $B$, $C$, and $D$ is split into sub-components, such that:
$A = A_1 \cdot A_2 \cdot A_3$, $B = B_1 \cdot B_2 \cdot B_3$, $C = C_1 \cdot C_2$, $D = D_1 \cdot D_2 \,\, ,$
where each sub-component corresponds to a variable in the original equation. The complete equation is given as follows. 

\begin{align*}
A_1 &= \frac{u(\phi) \left(a^2 - \frac{2M}{u(\phi)} + Q^2 + \frac{1}{u(\phi)^2}\right)}{\left(u(\phi) (a k - h)\left(Q^2 u(\phi) - 2M\right) - h\right)^3}, \\
A_2 &= \frac{ \left(\left(-\left(a^2 + Q^2\right) u(\phi)^2 + 2M u(\phi) - 1\right)\right)}{\left(u(\phi) (a k - h)\left(Q^2 u(\phi) - 2M\right) - h\right)^3}, \\[2ex]
A_3 &= \frac{ \left(u(\phi)(ak - h)\left(2M - Q^2 u(\phi)\right) + h\right)}{\left(u(\phi) (a k - h)\left(Q^2 u(\phi) - 2M\right) - h\right)^3}, \\[2ex]
B_1 &= \frac{2 \left(\left(a^2 + Q^2\right) u(\phi)^2 - 2M u(\phi) + 1\right)}{\left(u(\phi) (a k - h)\left(Q^2 u(\phi) - 2M\right) - h\right)^3}, \\[2ex]
B_2 &= \frac{ (M u(\phi)(ak - h) + h)}{\left(u(\phi) (a k - h)\left(Q^2 u(\phi) - 2M\right) - h\right)^3}, \\[2ex]
B_3 &= \frac{ (u(\phi) \left(u(\phi)\left(a^2\left(-(k^2 - 1)\right) + u(\phi)(h - ak)^2 \left(Q^2 u(\phi) - 2M\right) + h^2 + Q^2\right)\right)}{\left(u(\phi) (a k - h)\left(Q^2 u(\phi) - 2M\right) - h\right)^3}, \\[2ex]
C_1 &= \frac{-2 (1 - M u(\phi))}{\left(u(\phi) (a k - h)\left(Q^2 u(\phi) - 2M\right) - h\right)^3}, \\[2ex]
C_2 &= \frac{ \left(u(\phi)(ak - h)\left(2M - Q^2 u(\phi)\right) + h\right)}{\left(u(\phi) (a k - h)\left(Q^2 u(\phi) - 2M\right) - h\right)^3}, \\[2ex]
D_1 &= \frac{2 \left(\left(a^2 + Q^2\right) u(\phi)^2 - 2M u(\phi) + 1\right)}{\left(u(\phi) (a k - h)\left(Q^2 u(\phi) - 2M\right) - h\right)^3}, \\[2ex]
D_2 &= \frac{ \left(u(\phi) \left(u(\phi)\left(a^2\left(-(k^2 - 1)\right) + u(\phi)(h - ak)^2 \left(Q^2 u(\phi) - 2M\right) + h^2 + Q^2\right) - 2M\right) - k^2 + 1\right)}{\left(u(\phi) (a k - h)\left(Q^2 u(\phi) - 2M\right) - h\right)^3}
    \label{kn_orbit1}
\end{align*}
\subsection[\appendixname~\thesubsection]{Zeroth and First Order Terms in Kerr-Newman m-plot}
Here, we write the function $g$ representing the powers of $p$ in the zeroth order equation for the $m$-plot for the Kerr-Newman solution.
\begin{align*}
    g_7(L, a, e, M, q) &= 2 a e^3 M^8 - 2 a e M^8 - L M^8 \,\, , \\
    g_6(L, a, e, M, q) &= -3 a^2 e^2 L M^6 + 3 a^2 L M^6 - 2 a e^3 M^6 q^2 + 2 a e M^8 \nonumber \\
    & \quad + 2 a e M^6 q^2 + L^3 M^6 + 4 L M^8 + L M^6 q^2 \,\, , \\
    g_5(L, a, e, M, q) &= -2 a^3 e^3 M^6 + 2 a^3 e M^6 + 3 a^2 e^2 L M^6 - 11 a^2 L M^6 \nonumber \\
    & \quad + 6 a e L^2 M^6 + 4 a e M^8 - 3 a e M^6 q^2 - 7 L^3 M^6 \nonumber \\
    & \quad - 4 L M^8 - 6 L M^6 q^2 \,\, , \\
    g_4(L, a, e, M, q) &= -3 a^4 e^2 L M^4 + 3 a^4 L M^4 + 2 a^3 e^3 M^6 - a^3 e^3 M^4 q^2 \nonumber \\
    & \quad - 10 a^3 e M^6 + a^3 e M^4 q^2 + 12 a^2 e^2 L M^6 + 3 a^2 L^3 M^4 \nonumber \\
    & \quad + 10 a^2 L M^6 + 5 a^2 L M^4 q^2 - 30 a e L^2 M^6 - 3 a e L^2 M^4 q^2 \nonumber \\
    & \quad - 8 a e M^6 q^2 + a e M^4 q^4 + 16 L^3 M^6 + 4 L^3 M^4 q^2 \nonumber \\
    & \quad + 8 L M^6 q^2 + 2 L M^4 q^4 \,\, \\
    g_3(L, a, e, M, q) &= -4 a^5 e^3 M^4 + 4 a^5 e M^4 - 3 a^4 e^2 L M^4 - 4 a^4 L M^4 \nonumber \\
    & \quad + 12 a^3 e^3 M^6 - 3 a^3 e^3 M^4 q^2 + 18 a^3 e L^2 M^4 + 9 a^3 e M^4 q^2 \nonumber \\
    & \quad - 36 a^2 e^2 L M^6 - 12 a^2 e^2 L M^4 q^2 - 11 a^2 L^3 M^4 - 9 a^2 L M^4 q^2 \nonumber \\
    & \quad + 36 a e L^2 M^6 + 33 a e L^2 M^4 q^2 + 5 a e M^4 q^4 - 12 L^3 M^6 \nonumber \\
    & \quad - 18 L^3 M^4 q^2 - 5 L M^4 q^4 \,\, , \\
    g_2(L, a, e, M, q) &= -10 a^5 e^3 M^4 + a^5 e^3 M^2 q^2 - a^5 e M^2 q^2 + 30 a^4 e^2 L M^4 \nonumber \\
    & \quad + 3 a^4 e^2 L M^2 q^2 + a^4 L M^2 q^2 - 20 a^3 e^3 M^4 q^2 + a^3 e^3 M^2 q^4 \nonumber \\
    & \quad - 30 a^3 e L^2 M^4 - 9 a^3 e L^2 M^2 q^2 - 2 a^3 e M^2 q^4 + 60 a^2 e^2 L M^4 q^2 \nonumber \\
    & \quad + 3 a^2 e^2 L M^2 q^4 + 10 a^2 L^3 M^4 + 5 a^2 L^3 M^2 q^2 + 2 a^2 L M^2 q^4 \nonumber \\
    & \quad - 60 a e L^2 M^4 q^2 - 9 a e L^2 M^2 q^4 - a e M^2 q^6 + 20 L^3 M^4 q^2 \nonumber \\
    & \quad + 5 L^3 M^2 q^4 + L M^2 q^6 \,\, , \\
    g_1(L, a, e, M, q) &= 9 a^5 e^3 M^2 q^2 - 27 a^4 e^2 L M^2 q^2 + 11 a^3 e^3 M^2 q^4 + 27 a^3 e L^2 M^2 q^2 \nonumber \\
    & \quad - 33 a^2 e^2 L M^2 q^4 - 9 a^2 L^3 M^2 q^2 + 33 a e L^2 M^2 q^4 - 11 L^3 M^2 q^4 \,\, , \\
    g_0(L, a, e, M, q) &= -2 a^5 e^3 q^4 + 6 a^4 e^2 L q^4 - 2 a^3 e^3 q^6 - 6 a^3 e L^2 q^4 \nonumber \\
    & \quad + 6 a^2 e^2 L q^6 + 2 a^2 L^3 q^4 - 6 a e L^2 q^6 + 2 L^3 q^6 \,\, .
\end{align*}
Now, we write the function $f$ representing the powers of $p$ in the first-order equation for the $m$-plot for the Kerr-Newman solution.
\begin{align*}
    f_{10}(a, L, e, M, q) &= q^{2}L^{2} - 3a^{2}L^{2}e^{2} + 3a^{2}L^{2} - 12a^{2}e^{4} \nonumber \\
    &\quad + 12 a^{2} e^{2} + 2aLe{\left(-q^{2} e^{2} + q^{2} + 4 e^{2} L^{4}\right)} \,\, , \nonumber\\
    f_9(a, L, e, M, q) &= 20q^{2}a^{2}e^{4} - 20q^{2}a^{2}e^{2} - 8q^{2} L^{2} + 12a^{3}Le{\left(e^{2}-1\right)} \nonumber \\
    &\quad + 6a^{2}L^{2}e^{2} - 24a^{2}L^{2} + 16a^{2}e^{4} - 24a^{2}e^{2} \nonumber \\
    &\quad - 8 a L e {\left(q^{2} e^{2} + q^{2} - L^{2} + 2 e^{2} \right)} - 14 L^{4} \,\, , \nonumber \\
    f_8(a, L, e, M, q) &= -10 q^{4} a^{2} e^{4} + 10 q^{4} a^{2} e^{2} + 4 q^{4} L^{2} - 24 q^{2} a^{3} L e {\left(e^{2} - 1 \right)} \nonumber \\
    &\quad + 6 q^{2} a^{2} L^{2} e^{2} + 12 q^{2} a^{2} L^{2} - 32 q^{2} a^{2} e^{4} + 48 q^{2} a^{2} e^{2} \nonumber \\
    &\quad + 10 q^{2} L^{4} + 24 q^{2} L^{2} - 18 a^{4} L^{2} e^{2} + 18 a^{4} L^{2} \nonumber \\
    &\quad + 12 a^{2} L^{4} + 24 a^{2} L^{2} e^{2} + 72 a^{2} L^{2} + 4 a L e {\left(q^{4} e^{2} + q^{4} - q^{2} L^{2} \right)} \nonumber \\
    &\quad + 8 q^{2} e^{2} - 24 L^{2} + 72 L^{4} \,\, , \nonumber \\
    f_7(a, L, e, M, q) &= 24q^{4}a^{2}e^{4} - 36q^{4}a^{2}e^{2} - 24q^{4}L^{2} - 48q^{2}a^{2}L^{2}e^{2} \nonumber \\
    &\quad - 72q^{2}a^{2}L^{2} - 96q^{2}L^{4} - 32q^{2}L^{2} - 24a^{5}Le{\left(e^{2}-1\right)} \nonumber \\
    &\quad + 24a^{4}L^{2}e^{2} - 84a^{4}L^{2} + 8a^{3}Le{\left(3q^{2}e^{2}-6q^{2}+12L^{2}+4e^{2}\right)} \nonumber \\
    &\quad - 96a^{2}L^{4} - 240a^{2}L^{2}e^{2} - 96a^{2}L^{2} + 8aLe{\left(-3q^{4}e^{2}+15q^{2}L^{2}+4q^{2} \right)} \nonumber \\
    &\quad + 48 L^{2} - 176 L^{4} \,\, , \nonumber \\
    f_6(a, L, e, M, q) &= -6q^{6}a^{2}e^{4} + 9q^{6}a^{2}e^{2} + 6q^{6}L^{2} - 9q^{4}a^{4}e^{4} \nonumber \\
    &\quad + 9q^{4}a^{4}e^{2} + 15q^{4}a^{2}L^{2}e^{2} + 18q^{4}a^{2}L^{2} + 30q^{4}L^{4} \nonumber \\
    &\quad + 48q^{4}L^{2} + 18q^{2}a^{4}L^{2}e^{2} + 27q^{2}a^{4}L^{2} + 12q^{2}a^{4}e^{4} \nonumber \\
    &\quad - 24q^{2}a^{4}e^{2} + 48q^{2}a^{2}L^{4} + 444q^{2}a^{2}L^{2}e^{2} + 144q^{2}a^{2}L^{2} \nonumber \\
    &\quad + 16q^{2}a^{2}e^{2} + 336q^{2}L^{4} + 16q^{2}L^{2} - 15a^{6}L^{2}e^{2} + 15a^{6}L^{2} \nonumber \\
    &\quad - 12a^{6}e^{4} + 12a^{6}e^{2} - 6a^{5}Le{\left(3q^{2}e^{2}-3q^{2}-4e^{2}+24\right)} \nonumber \\
    &\quad + 15a^{4}L^{4} + 276a^{4}L^{2}e^{2} + 132a^{4}L^{2} + 16a^{4}e^{4} \nonumber \\
    &\quad + 48a^{4}e^{2} - 2a^{3}Le{\left(-9q^{4}+24q^{2}L^{2}+36q^{2}e^{2}\right)} \nonumber \\
    &\quad + 60q^{2} + 288L^{2} + 128e^{2} + 48 + 288a^{2}L^{4} + 672a^{2}L^{2}e^{2} \nonumber \\
    &\quad + 48a^{2}L^{2} - 2aLe{\left(-3q^{6}e^{2}+18q^{4}L^{2}+24q^{4}\right)} \nonumber \\
    &\quad + 360q^{2}L^{2} + 16q^{2} + 320L^{2} + 208L^{4} \,\, , \nonumber \\ 
    f_5(a, L, e, M, q) &= -24q^{6}L^{2} - 6q^{4}a^{4}e^{4} + 12q^{4}a^{4}e^{2} - 258q^{4}a^{2}L^{2}e^{2} \nonumber \\
    &\quad - 72q^{4}a^{2}L^{2} - 32q^{4}a^{2}e^{2} - 204q^{4}L^{4} - 32q^{4}L^{2} \nonumber \\
    &\quad - 12q^{2}a^{6}e^{4} + 12q^{2}a^{6}e^{2} - 300q^{2}a^{4}L^{2}e^{2} - 84q^{2}a^{4}L^{2} \nonumber \\
    &\quad - 32q^{2}a^{4}e^{4} - 96q^{2}a^{4}e^{2} - 288q^{2}a^{2}L^{4} - 1632q^{2}a^{2}L^{2}e^{2} \nonumber \\
    &\quad - 96q^{2}a^{2}L^{2} - 512q^{2}L^{4} - 36a^{7}Le\left(e^{2} - 1\right) - 6a^{6}L^{2}e^{2} \nonumber \\
    &\quad - 36a^{6}L^{2} - 72a^{6}e^{2} + 24a^{5}Le\left(q^{2}e^{2} + 3q^{2} + 5L^{2} + 16e^{2} + 6\right) \nonumber \\
    &\quad - 78a^{4}L^{4} - 1152a^{4}L^{2}e^{2} - 72a^{4}L^{2} - 96a^{4}e^{4} \nonumber \\
    &\quad + 4a^{3}Le\big(9q^{4}e^{2} + 15q^{4} + 144q^{2}L^{2} + 152q^{2}e^{2}\big) \nonumber \\
    &\quad + 48q^{2} + 288L^{2} + 96e^{2}\big) - 384a^{2}L^{4} - 576a^{2}L^{2}e^{2} \nonumber \\
    &\quad + 8aLe\big(3q^{6} + 54q^{4}L^{2} + 8q^{4} + 196q^{2}L^{2} + 48L^{2}\big) \nonumber \\
    &\quad - 96L^{4} \,\, , \nonumber \\
\end{align*}
\begin{align*}
    f_4(a, L, e, M, q) &= -4q^{8}aLe + 4q^{8}L^{2} + 48q^{6}a^{2}L^{2}e^{2} + 12q^{6}a^{2}L^{2} \nonumber \\
    &\quad + 24q^{6}a^{2}e^{2} - 84q^{6}aL^{3}e - 48q^{6}aLe + 40q^{6}L^{4} + 24q^{6}L^{2} \nonumber \\
    &\quad + 72q^{4}a^{4}L^{2}e^{2} + 12q^{4}a^{4}L^{2} + 24q^{4}a^{4}e^{4} + 72q^{4}a^{4}e^{2} \nonumber \\
    &\quad + 72q^{4}a^{2}L^{4} + 1440q^{4}a^{2}L^{2}e^{2} + 72q^{4}a^{2}L^{2} - 1392q^{4}aL^{3}e \nonumber \\
    &\quad + 456q^{4}L^{4} + 24q^{2}a^{6}L^{2}e^{2} + 4q^{2}a^{6}L^{2} + 72q^{2}a^{6}e^{2} + 32q^{2}a^{4}L^{4} \nonumber \\
    &\quad + 1728q^{2}a^{4}L^{2}e^{2} + 72q^{2}a^{4}L^{2} + 288q^{2}a^{4}e^{4} + 576q^{2}a^{2}L^{4} \nonumber \\
    &\quad + 1728q^{2}a^{2}L^{2}e^{2} - 1152q^{2}aL^{3}e + 288q^{2}L^{4} - 24a^{8}e^{4} + 24a^{8}e^{2} \nonumber \\
    &\quad - 4a^{7}Le\big(-q^{2}e^{2} + q^{2} + 16e^{2} + 12\big) + 336a^{6}L^{2}e^{2} + 24a^{6}L^{2} \nonumber \\
    &\quad + 192a^{6}e^{4} - 12a^{5}Le\big(q^{4} + 5q^{2}L^{2} + 48q^{2}e^{2} + 12q^{2} + 32L^{2} + 64e^{2}\big) \nonumber \\
    &\quad + 136a^{4}L^{4} + 1152a^{4}L^{2}e^{2} - 4a^{3}Le\big(q^{6}e^{2} + 3q^{6} + 36q^{4}L^{2} \nonumber \\
    &\quad + 132q^{4}e^{2} + 36q^{4} + 432q^{2}L^{2} + 288q^{2}e^{2} + 192L^{2}\big) + 92a^{2}L^{4} \,\, , \nonumber \\
    f_3(a, L, e, M, q) &= -8q^{8}a^{2}e^{2} + 16q^{8}aLe - 8q^{8}L^{2} - 8q^{6}a^{4}e^{4} \nonumber \\
    &\quad - 24q^{6}a^{4}e^{2} + 16q^{6}a^{3}Le^{3} + 48q^{6}a^{3}Le - 184q^{6}a^{2}L^{2}e^{2} \nonumber \\
    &\quad - 24q^{6}a^{2}L^{2} + 352q^{6}aL^{3}e - 176q^{6}L^{4} - 24q^{4}a^{6}e^{2} + 48q^{4}a^{5}Le \nonumber \\
    &\quad - 288q^{4}a^{4}L^{2}e^{2} - 24q^{4}a^{4}L^{2} - 336q^{4}a^{4}e^{4} + 576q^{4}a^{3}L^{3}e \nonumber \\
    &\quad + 672q^{4}a^{3}Le^{3} + 8q^{4}a^{3}Le^{3}\left(23q^{2} + 84\right) - 288q^{4}a^{2}L^{4} \nonumber \\
    &\quad - 672q^{4}a^{2}L^{2}e^{2} - 16q^{4}a^{2}L^{2}e^{2}\left(23q^{2} + 84\right) + 672q^{4}aL^{3}e \nonumber \\
    &\quad + 8q^{4}aL^{3}e\left(23q^{2} + 84\right) - 336q^{4}L^{4} + 8q^{2}a^{8}e^{4} - 8q^{2}a^{8}e^{2} \nonumber \\
    &\quad - 16q^{2}a^{7}Le^{3} + 16q^{2}a^{7}Le - 104q^{2}a^{6}L^{2}e^{2} - 8q^{2}a^{6}L^{2} \nonumber \\
    &\quad - 384q^{2}a^{6}e^{4} + 224q^{2}a^{5}L^{3}e + 768q^{2}a^{5}Le^{3} + 96q^{2}a^{5}Le^{2}e\left(3q^{2} + 8\right) \nonumber \\
    &\quad - 112q^{2}a^{4}L^{4} - 768q^{2}a^{4}L^{2}e^{2} - 192q^{2}a^{4}L^{2}e^{2}\left(3q^{2} + 8\right) \nonumber \\
    &\quad + 768q^{2}a^{3}L^{3}e + 96q^{2}a^{3}L^{3}e\left(3q^{2} + 8\right) - 384q^{2}a^{2}L^{4} \nonumber \\
    &\quad - 80a^{8}e^{4} + 160a^{7}Le^{3} + 8a^{7}Le^{3}\left(13q^{2} + 20\right) - 160a^{6}L^{2}e^{2} \nonumber \\
    &\quad - 16a^{6}L^{2}e^{2}\left(13q^{2} + 20\right) + 160a^{5}L^{3}e + 8a^{5}L^{3}e\left(13q^{2} + 20\right) \nonumber \\
    &\quad - 80a^{4}L^{4} \,\, , \nonumber \\
    f_2(a, L, e, M, q) &= q^{10}a^{2}e^{2} - 2q^{10}aLe + q^{10}L^{2} + q^{8}a^{4}e^{4} \nonumber \\
    &\quad + 3q^{8}a^{4}e^{2} - 2q^{8}a^{3}Le^{3} - 6q^{8}a^{3}Le + 26q^{8}a^{2}L^{2}e^{2} \nonumber \\
    &\quad + 3q^{8}a^{2}L^{2} - 50q^{8}aL^{3}e + 25q^{8}L^{4} + 3q^{6}a^{6}e^{2} - 6q^{6}a^{5}Le \nonumber \\
    &\quad + 48q^{6}a^{4}L^{2}e^{2} + 3q^{6}a^{4}L^{2} + 192q^{6}a^{4}e^{4} - 96q^{6}a^{3}L^{3}e \nonumber \\
    &\quad - 384q^{6}a^{3}Le^{3} - 2q^{6}a^{3}Le^{3}\left(13q^{2} + 192\right) + 48q^{6}a^{2}L^{4} \nonumber \\
    &\quad + 384q^{6}a^{2}L^{2}e^{2} + 4q^{6}a^{2}L^{2}e^{2}\left(13q^{2} + 192\right) - 384q^{6}aL^{3}e \nonumber \\
    &\quad - 2q^{6}aL^{3}e\left(13q^{2} + 192\right) + 192q^{6}L^{4} - q^{4}a^{8}e^{4} + q^{4}a^{8}e^{2} \nonumber \\
    &\quad + 2q^{4}a^{7}Le^{3} - 2q^{4}a^{7}Le + 22q^{4}a^{6}L^{2}e^{2} + q^{4}a^{6}L^{2} \nonumber \\
    &\quad + 288q^{4}a^{6}e^{4} - 46q^{4}a^{5}L^{3}e - 576q^{4}a^{5}Le^{3} - 48q^{4}a^{5}Le^{3}\left(q^{2} + 12\right) \nonumber \\
    &\quad + 23q^{4}a^{4}L^{4} + 576q^{4}a^{4}L^{2}e^{2} + 96q^{4}a^{4}L^{2}e^{2}\left(q^{2} + 12\right) \nonumber \\
    &\quad - 576q^{4}a^{3}L^{3}e - 48q^{4}a^{3}L^{3}e\left(q^{2} + 12\right) + 288q^{4}a^{2}L^{4} \nonumber \\
    &\quad + 100q^{2}a^{8}e^{4} - 200q^{2}a^{7}Le^{3} - 2q^{2}a^{7}Le^{2}e\left(11q^{2} + 100\right) \nonumber \\
    &\quad + 200q^{2}a^{6}L^{2}e^{2} + 4q^{2}a^{6}L^{2}e^{2}\left(11q^{2} + 100\right) - 200q^{2}a^{5}L^{3}e \nonumber \\
    &\quad - 2q^{2}a^{5}L^{3}e\left(11q^{2} + 100\right) + 100q^{2}a^{4}L^{4} \,\, , \nonumber \\
\end{align*}
\begin{align*}
    f_1(a, L, e, M, q) &= -54q^{8}a^{4}e^{4} + 216q^{8}a^{3}Le^{3} - 324q^{8}a^{2}L^{2}e^{2} + 216q^{8}aL^{3}e \nonumber \\
    &\quad - 54q^{8}L^{4} - 96q^{6}a^{6}e^{4} + 384q^{6}a^{5}Le^{3} - 576q^{6}a^{4}L^{2}e^{2} + 384q^{6}a^{3}L^{3}e \nonumber \\
    &\quad - 96q^{6}a^{2}L^{4} - 42q^{4}a^{8}e^{4} + 168q^{4}a^{7}Le^{3} - 252q^{4}a^{6}L^{2}e^{2} + 168q^{4}a^{5}L^{3}e \nonumber \\
    &\quad - 42q^{4}a^{4}L^{4} \,\, , \nonumber \\
    f_0(a, L, e, M, q) &= 6q^{10}a^{4}e^{4} - 24q^{10}a^{3}Le^{3} + 36q^{10}a^{2}L^{2}e^{2} - 24q^{10}aL^{3}e \nonumber \\
    &\quad + 6q^{10}L^{4} + 12q^{8}a^{6}e^{4} - 48q^{8}a^{5}Le^{3} + 72q^{8}a^{4}L^{2}e^{2} - 48q^{8}a^{3}L^{3}e \nonumber \\
    &\quad + 12q^{8}a^{2}L^{4} + 6q^{6}a^{8}e^{4} - 24q^{6}a^{7}Le^{3} + 36q^{6}a^{6}L^{2}e^{2} - 24q^{6}a^{5}L^{3}e \nonumber \\
    &\quad + 6q^{6}a^{4}L^{4} \nonumber \,\, .
\end{align*}

\twocolumngrid

\nocite{*}
\bibliographystyle{apsrev4-2}
\bibliography{references}

\end{document}